\documentclass{elsarticle}
\usepackage[utf8]{inputenc}
\usepackage{amsmath}
\usepackage{algorithmicx}
\usepackage{algpseudocode}
\usepackage{algorithm}
\usepackage{multirow}
\usepackage{balance}
\usepackage{graphics}
\usepackage{times}
\usepackage{balance}
\usepackage{graphicx}
\usepackage{subfig}
\usepackage{epstopdf}
\usepackage{rotating}
\usepackage{xspace}
\usepackage{enumitem}
\usepackage{eso-pic} 
\usepackage{booktabs}
\usepackage{colortbl}

\graphicspath{{./figures/}}
\DeclareGraphicsExtensions{.pdf,.jpeg,.png,.jpg}

\begin{document} 
\begin{frontmatter}

\title{An edge-fog-cloud platform for anticipatory  learning process designed for Internet of Mobile Things
}
\author[label1]{Hung Cao}
\fntext[unb]{Email: \texttt{\{hcao3,monicaw\}@unb.ca}.}
\author[label1]{Monica Wachowicz}
\author[label2]{Chiara Renso}
\fntext[isti]{Email: \texttt{\{chiara.renso,emanuele.carlini\}@isti.cnr.it}.}
\author[label2]{Emanuele Carlini}
\address[label1]{People in Motion Lab, University of New Brunswick, Fredericton, Canada}
\address[label2]{HPC-Lab, ISTI-CNR, Via Moruzzi 1, Pisa, Italy}

%\author{Hung Cao\fnref{unb}}
 %\fntext[unb]{University of New Brunswick. email: \texttt{XXXX}.}
%\author{Monica Wachowicz\fnref{unb}} 
%\author{Chiara Renso\fnref{isti}}\fntext[isti]{ISTI-CNR, Pisa, Italy. email: \texttt{\{name.surname\}@isti.cnr.it}.}
%\author{Emanuele Carlini\fnref{isti}}

\begin{abstract}
This paper presents a novel architecture for data analytics targeting an anticipatory learning process in the context of the Internet of Mobile Things. The architecture is geo-distributed and composed by edge, fog, and cloud resources that operate collectively to support such an anticipatory learning process. We designed the architecture to manage large volumes of data streams coming from the IoMT devices, analyze in successive phases climbing up in the hierarchy of resources from edge, fog and cloud. We discuss the characteristics of the analytical tasks at each layer. We notice that the amount of data being transported in the network decreases going from the edge, to the fog and finally to the cloud, while the complexity of the computation increases. Such design allows to support different kind of analytical needs, from real-time to historical according to the type of resource being utilized. We have implemented the proposed architecture as a proof-of-concept using the transit data feeds from the area of Greater Moncton, Canada. 
\end{abstract}


\end{frontmatter}

\section{Introduction}
\label{sec:introduction}
The \textit{“Things”} in the Internet of Things (IoT) are usually \textit{“smart devices”} which can sense the surroundings around their location and interact among themselves without human intervention \cite{lopez2012adding}. 
%They can be traffic lights, parking meters, electric vehicles, smartphones, and wearable devices \cite{}.  Atzori et al. \cite{} regards the IoT paradigm as a result of the convergence between the network and semantics, making IoT devices capable of extending human senses. From a network perspective, the research has been focussed on the architecture components of IoT ranging from edge computing and fog computing, to combining short- and long range networks, and developing standards [22–27]. From a semantic perspective, many applications have been developed in the fields of smart factory [9], smart home [10], and smart health care [11]. Tsai, Chun-Wei, et al.[12] provide a brief review of data analytics for IoT including  classification, clustering, association rules and sequential patterns [12].
In this paper, we are particularly interested in the Internet of Moving Things (IoMT) because of its potential to bridge the gap between an IoT device, its environment and a user. As a matter of fact, \textit{“mobile smart devices”} are integrated with everyday life of users, becoming personal, timely and relevant to use. In most IoMT applications, mobility and the geo-distribution of IoMT devices play an important role in how much the mobile smart devices are interacting; demanding a rich scenario of communication among themselves such as V2V (vehicle-to-vehicle), V2I (vehicle-to-infrastructure), and V2H (vehicle-to-home) \citep{gerla2014internet,lu2014connected}. 

IoMT presents a scenario in which unbounded streams of data are arriving at the sensing devices (i.e. sensors installed on a traveling vehicle or a personal device) as a high rate data stream. These data have to be processed \textit{“on the fly”} to detect anomalies, operational exceptions, deliver real-time alerts, and trigger automated actions. The real time actions to be taken based on the received and analyzed data is referred to an an Anticipatory Learning System. An anticipatory system is defined by Rosen \cite{rosen2012anticipatory} as \textit{“a system whose current state is determined by a (predicted) future state”} while Nadin \cite{nadin2010anticipatory} has previously defined it as \textit{“a system whose current state is determined not only by a past state, but also by possible future states.”} Anticipatory systems are therefore different from predictive systems as the formed describes an entire process that include prediction but also a feedback to the user to change their behavior according to the prediction model. In today's interconnected world, many of the devices that populate the Internet of Things are equipped with enough sensors and computational power to play a role in anticipatory systems. In fact, Pejovic and Musolesi \cite{pejovic2015anticipatory} acknowledged that the further proliferation of real-world anticipatory systems will rely on the deployment of IoT, and in particular IoMT. The unprecedented amount of data streams generated at the network edge by sensors and devices and the emergence of applications requiring low latency and real time actions calls for geo-distribution processing environment in contrast to the centralized environments. Mainly because IoMT devices will be able to seamlessly interact with the environment and sense feedback which will guide an anticipatory learning process \cite{pejovic2015anticipatory}.  But in order to support such an anticipatory learning process, IoMT requires a vast heterogeneity of resources from cloud-to-things. ranging from edge, fog and cloud computing resources. In fact, IoMT paired with Edge-Fog-Cloud resources gives an unprecedented opportunity to collect large amount of streaming real time data from the mobile devices and perform computations at all levels. 

For example, edge network devices may be located at a moving vehicle and thus having an active role in communicating and delivering high quality streaming to and from moving vehicles through APs positioned along in the environment such as highways \cite{stojmenovic2014fog}. Another example is a smart camera in a traffic management system that can produce a few gigabits per second. On one hand, streaming this data to the centre to process and get back the response not only requires bandwidth but it also may cause delays to get the actionable knowledge. On the other hand, it is not a realistic assumption that edge devices (i.e. cameras) have the necessary computation power and amount of information to process the data in a timely and correct fashion. Therefore, processing, storing, and analysing videos on the fog node of a network may be a sustainable solution. A well-designed Edge-Fog-Cloud architecture will allow us to support immediate feedback to the users based on timely predictions (e.g. traffic), rather than moving the data to data centers.  

The next generation of anticipatory applications based on the integration of Edge, Fog, and Cloud computing using IoMT devices are expected to support the following capabilities: 

\begin{itemize}
\item[--] \textbf{Scalability:} anticipatory applications will not require intensive analytics all the time; thus collecting, processing, and analyzing the raw data close to the source IoMT devices can produce momentary relevant content that can be used as a basis for intelligent decision making. 
\item[--] \textbf{Mobility and Geo-distribution:} in most anticipatory applications, mobility and geo-distribution are essential for supporting real-time analytics. 
\item[--] \textbf{Low Latency:} for many anticipatory applications, the edge analytics can identify actions in real time and avoid delays between the sensor-registered event and the reaction to that event. 
\end{itemize}

Towards addressing these challenges, this paper proposes an IoMT computing architecture for supporting an anticipatory learning process which can allow sensing the context of a surrounding environment of an IoMT device as well as from one location to another in real-time. Our research premise is that IoMT data arrives at the Edge, Fog and/or Cloud and requires processing at various speeds (e.g. stream versus batch, near-real-time versus real-time). Specifically, the architecture we propose consists of an edge layer, a fog layer and cloud layer which contain computational nodes that are needed for all phases of an anticipatory learning process and specifically in data analytics tasks. In our approach, the data analytics tasks are distributed anywhere through the combination of Edge, Fog, and Cloud resources with the objective of reducing the unpredictable network latency, expensive bandwidth, resource-prohibitive and location-awareness concerns of the Internet of Mobile Things. 

We demonstrate the usefulness of this platform with a case study run on a real dataset of bus traces collected by the CODIAC Transit agency in the Greater Moncton area (Canada). In these experiments we focus on the predictive task of an anticipatory process and how the predictive learning process and show examples of the tasks to be implemented in the proposed architecture at the different layers. We discuss the analytic tasks at the Edge, Fog and Cloud according to the anticipatory learning process. 

The rest of the paper is organized as follows. Section \ref{sec:related}  discusses the related works, while the proposed anticipatory learning process is described in Section \ref{sec:alp} and the platform for anticipatory learning in IoMT  in described in Section \ref{sec:platform}. Section \ref{sec:casestudy} reports the case study on transit data. Finally, conclusions and future research are given in Section 6.

\section{Related Work}
\label{sec:related}
A fairly systematic overview of IoT was recently published in \cite{li2015internet}, therefore in this section we mainly focuses on IoMT approaches. Overall,  IoMT devices are equipped with many kinds of sensors, ranging from accelerometers and gyroscopes to proximity, light, and ambient sensors, as well as microphones and cameras. It is indisputable that IoMT devices produce a large amount of heterogeneous data over time. This poses a challenge to capture, perception, management and processing within an acceptable time \cite{chen2014big}. The nature of IoMT data is usually multi-model, diverse, and heterogeneous, often streamed at high speed, and may be uncertain due to biases, noise and irregularities caused by latency, ambiguities, deception, or approximation. Different types of computing having been proposed for processing and analyzing IoMT data  (See Table \ref{table:related} for an overview). 

\textbf{Edge Computing}, according to Shi et al. \cite{shi2016edge}, refers to \textit{“the enabling technologies allowing computation to be performed at the edge of the network, on downstream data on behalf of cloud services and upstream data on behalf of IoT services.”} The rationale of edge computing is that data processing and analytics should happen close to IoMT devices. Very few attempts were found in the literature using edge computing for allowing data analytics near to the IoMT devices in spite of the recent statistics that 45\% of IoT data will be processed and analyzed at the edge of the network by 2019 \cite{shi2016edge}. 

\textbf{Mobile Edge Computing} was introduced by Nokia Networks and Intel \cite{intel2014} with the aim of supporting a base station as an intelligent service hub that can collect real-time network data such as cell congestion and subscriber locations. The ETSI Industry Specification Group (ISG) has defined Mobile Edge Computing as \textit{“a concept that provides an IT service environment and cloud-computing capabilities at the edge of the mobile network, within the Radio Access Network (RAN) and in close proximity to mobile subscribers.”} \cite{hu2015mobile}. 

\textbf{Fog computing} was first introduced by Cisco as a bridge between the Edge and the Cloud \cite{bonomi2012fog}. Other paradigms having a similar concept were also proposed in the literature such as Cloudlet \cite{satyanarayanan2009case} and Mobile Cloud Computing \cite{khan2013towards} as well as Mobile Edge Computing \cite{intel2014}. Typically, Bonomi et al \cite{bonomi2014fog} describe fog computing as a hierarchical distributed architecture on the edge of the network to process IoMT data with a low latency, location awareness, and mobility support. Several scenarios have been envisaged to apply fog computing, including Augmented Reality (AR), Real-time Video Analytics, Mobile Big Data Analytics, Smart Grid, Smart Traffic Lights and Connected Vehicles, Decentralized Smart Building Control, Wireless Sensors and Actuators Networks \cite{stojmenovic2014fog,yi2015survey,bonomi2014fog}. Unfortunately, none of these scenarios has been actually implemented so far.

\textbf{Mobile Cloud Computing} was proposed to overcome the shortage in computing power and storage capacity of mobile devices by leveraging the services of Cloud Computing to offload computation for these end devices \cite{ali2009green,kumar2010cloud}. Khan et al. \cite{khan2013towards} define Mobile Cloud Computing as \textit{“a service that allows resource-constrained mobile users to adaptively adjust processing and storage capabilities by transparently partitioning and offloading the computationally intensive and storage demanding jobs on traditional cloud resources by providing ubiquitous wireless access”}. In this paper, Fog Computing is used to encompass all these three paradigms (Mobile Edge Computing, Mobile Cloud Computing, Fog Computing). 

\textbf{Cloud Computing} has dominated the scalable infrastructures as well as processing engines developed to support SaaS, PaaS and IaaS models during the last decade, leading to a trend of Everything as a Service (XaaS) \cite{banerjee2011everything}. However, as Hong et al. \cite{hong2013mobile} point out, existing Cloud Computing models have been designed for traditional web applications rather than future IoMT applications running on various mobile and sensor devices. Moreover, clouds are far from the flawless utility computing model because their current network bandwidth and reliability are not adequate to support the short response time needed for processing the large quality of data generated by IoMT devices in real-time. Therefore, Edge Computing has emerged as a new computing platform for geo-spatially distributed, large-scale and latency-sensitive future IoMT applications.

Table \ref{table:related} illustrates the main research efforts towards developing different architectures for supporting IoMT and reveals the variety of IoMT devices currently being used for different applications. The data is being streamed in real time to the cloud but the processing is still batch in most of the cases. Our approach is the first one, to the best of our knowledge, where the streamed data from IoMT sensors are analyzed at the edge in real time, at the fog in near real time and eventually at the cloud as historical data. 
 
\renewcommand{\arraystretch}{1.3} 

 \begin{table}
 \small
 \centering 
\begin{tabular}{@{} m{6em} m{6em} m{6em} m{12em} m{3em} @{}} \toprule 
\bf{Type of Computing} &  \bf{IoMT devices} &  {\small \bf{Processing and Analytics}} &  \bf{Applications} &  \bf{Ref.}\\
 \toprule
Cloud & RFID tags, BLE & historical & Anticipatory Ubiquitous Computing
 & \cite{atzmueller2016sensing}\\
%\hline
Cloud &  WiFi, BLE &  historical &  Location/Future Movement Prediction 
&  \cite{nahrstedt2016internet}\\
%\hline
Cloud &  Spatial-Temporal Data, GPS, Camera, Environmental Sensors & 
historical &  Moving Object Map Analytics (MOMA), Contextual Spatial-Temporal Analytics  &  \cite{sun2016moving}\\
%\hline
Cloud & 
GPS, Rain Gauge Data, Road Incident Report, Social Media & 
mixed (historical and streaming) & 
Urban Trajectory Data Analytics System & 
\cite{vieira2015usapiens}\\
\hline
Edge + Cloud & BLE &  near-time &  O/D Transportation Planning & 
 \cite{herrera2016smart}\\
% \hline
Edge + Cloud &  RFID tags &  historical & 
RFID Ecosystem for management, IoT applications & \cite{welbourne2009building}\\
%\hline
 Edge + Cloud &  Sensors, traffic lights &  real-time & 
Virtual Object (VO) model to enrich context information with Cognitive Internet of Things &  \cite{somov2013supporting}\\
\hline
Edge &  Phone Camera &  real-time & 
Pedestrian Safety Detection (Offline Training/Online Detection)&
 \cite{wang2012walksafe}\\
 %\hline
 Edge & Sensors, RFID &  real-time & 
Proposed the Smart Object framework to encapsulate
RFID, sensor, Internet-based data & 
\cite{lopez2012adding}\\
%\hline
Edge & 
Wearable sensors, GPS receivers, Laptop, Smartphone & 
real-time & 
wearable system which 
can learn context-dependent personal preferences & 
\cite{krause2006context}\\
\bottomrule
\end{tabular}
\caption{Overview of approaches for edge, fog and cloud computation for IoMT applications }
\label{table:related}
\end{table}

All of the aforementioned research work described in Table \ref{table:related} has not been developed based on fog computing as well as multilayer architecture such as Edge-Fog-Cloud. Very few exceptions are found in the literature which includes the emerging wave of the Internet of Vehicles \citep{gerla2014internet,lu2014connected} where Fog Computing is expected to play an active role in delivering high speed data streams to moving vehicles through APs positioned along highways \cite{stojmenovic2014fog}. In \cite{hong2013mobile}, Hong et al. propose a mobile fog architecture as a spatio-temporal event processing system that uses continuous querying for traffic monitoring and distributed complex event processing system using vehicle tracking cameras. 
Pejovic and Mulosi \cite{pejovic2015anticipatory} provides a classification of machine learning techniques for an anticipatory learning process that have been proved to work in the domains of speech recognition, place categorization, location awareness and call prediction. The techniques vary from threshold-based learning, tree based learner to hidden Markov models, and Bayesian networks.  The anticipatory computing architecture is usually deployed in the Cloud, and it supports batch processing of data gathered from “always-on” devices. But in our case, real-world anticipatory systems will have to rely on a multi-layered computing platform for processing and analyzing a vast amount of data streams generated by IoMT devices in real-time. Mainly because it will not be possible to transport the data streams to the cloud and provide any anticipatory feedback that is tightly integrated with personalized patterns and actions of users. Flexible fog nodes in terms of hardware and software architecture are important to respond to IoMT needs than a cloud cluster is. Fog nodes will be the catalyst of IoMT and connect the ecosystem which is needed for real-world anticipatory systems. Our proposed anticipatory behaviour platform is described in more detail in the following section.

\section{The Anticipatory Learning Process}
\label{sec:alp}
An anticipatory system can be broadly defined as a system that changes its state according to the prediction performed on a model of itself and the surrounding environment \cite{rosen2012anticipatory}.  Clearly, in anticipatory systems the ability to correctly predict the future context plays a pivotal role. In the context of IoMT, prediction is an hard challenge due to the complexity of the environment and the rapidly changing situations. An anticipatory learning process deals with such dynamicity: not only it is necessary a proper sensing of the environment to instruct the current prediction models, but also there is the need to tune the prediction models such that to adapt them to the surrounding context.
From a modeling point of view, the anticipatory learning process for IoMT consists in five main phases: sensing, data preprocessing, contextualization, prediction and feedback.

\subsection{Sensing} 
In this first phase, the aim is to listen to the environment for gathering a multitude of physical signals that are descriptive of the environment. In the context of IoMT, this is made possible by the today’s devices complex hardware and their extreme ubiquity. It is often the case in which an IoMT device is equipped with hardware that can capture sound, images, temperature, positions, light and many other features of the  surrounding environment.  From an anticipatory system perspective, IoMT devices are capable of extending human senses and communicate without human intervention. 

\subsection{Data preprocessing }
Data preprocessing usually refers to cleaning, filtering, aggregating, wrangling, or dealing with missing data. It is important to point out that the right choice of preprocessing techniques will have a positive impact on the ability to predict and seek for intelligent actions within an anticipatory learning process. Data preprocessing can consists of the following tasks:
\begin{itemize}
\item[--] Dealing with missing data: there are several ways to tackle this problem. For large data sets, deleting samples based on missing values is expected not to be a problem. But for a small data sets, having a large number of missing values may affect our later computing stages. In this case, missing values can be replaced based on predictive models \cite{larose2014discovering}. 
\item[--] Filtering: IoMT devices may produce error and noise data; in order to minimize the impact of errors on input data on succeeding analyses, the process of defining, detecting and correcting errors in given data can be applied. Some new approaches can be found in \cite{saez2016inffc}.
\item[--] Aggregating: in some anticipatory applications the summary form of data could be enough for statistical analysis \cite{barnaghi2013data}. Aggregation may be applied to diminish the bandwidth consumption as well as the burden on system.  
\item[--] Cleansing: IoMT data sometimes comes with irrelevant or inaccurate parts of data. Cleansing brings some major advantages such as reducing computational time and complexity due to fewer data features, improving the performance of the predictive model \cite{kuhn2013applied}.  
\item[--] Wrangling: in some applications, the raw data may need to be converted or changed into another format in order to make it more convenient for the later computing stages.

\end{itemize}

\subsection{Data Contextualization} 

Data contextualization plays an important role in an anticipatory learning process because it helps to explain a phenomenon, detect and better understand an abnormal behavior \cite{perera2014context}. It can reinforce different perspectives and may have different dimensions such as geographical, physical, social, temporal. Perera et al. \cite{perera2014context} categorise context into two main types: primary and secondary context, meanwhile Van Bunningen et al. \cite{van2005context} propose two broader categories: operational and conceptual. Others have distinguished different contexts based on location, identity, activity, time \cite{abowd1999towards} as well as based on the categorisation technique – sensed, static, profiled, derived \cite{henricksen2003framework}.
The proposed learning process is designed to support severals automated tasks for data contextualization. These tasks will depend on the type application (see Section \ref{sec:casestudy} for an example of a  set of tasks designed for a smart transit application). But in any application, context is used to enrich the raw data gathered by IoMT devices with information that is needed for the anticipatory learning process. Moreover, the automated execution of all the tasks of the data contextualization phase task is crucial in our platform, especially because of the IoMT devices where inter-device communication needs to be established quickly and efficiently and actions must be coordinated together. Finally, the raw data streams are enriched by tagging semantic attributes (e.g. place, social and activity attributes) for later analysis.

\subsection{Prediction} 
In anticipatory computing, context prediction and intelligent-driven action play the major roles to assist a system to change its behavior based on their prediction. A predictive model that provides information about possible future states of the surrounding environment is the key component to do this.
One point in prediction is the presence of both labeled data and unlabeled data on which to apply  supervised or unsupervised prediction techniques. Supervised techniques use labeled data with the correct answers and trained them to learn a model that is then applied to new data to produce predictions or classifications. Unsupervised techniques, in contrast, use only unlabeled data to find behavioral patterns and attempt can be done to predict common patterns.
In our case, models are learned from the cleaned and contextualized data to predict the behavior of the observed IoMT objects in the future. An example is the prediction of the speed of a vehicle in a future time instant by learning a model on the historical speed data. Prediction is typically performed by machine learning algorithms like Linear Regression, SVN, Random Forest or Gradient Boost to name a few.

\subsection{Feedback}
The feedback is a relevant result of the analytical tasks that will guide the actions to be done by users of an anticipatory learning process. Feedback is sent back from any analytical layer (i.e. edge, fog, cloud) to IoMT actuators to take immediate actions. Feedback can be real time, near real time or historical depending on the layer where it is computed.  Examples of real time feedback are the detected anomalies on the operational behavior of the device at the edge, or abnormal behavior in the movement of a traveling object detected at the fog or the cloud.

\section{The Anticipatory Behavioral Platform}
\label{sec:platform}
Our proposal is to map our anticipatory learning process in the context of IoMT using a multilayer architecture, i.e. an edge-fog-cloud architecture.  The aim is to exploit the combination of different computation resources available at the edge nodes, fog nodes and cloud clusters in order to provide meaningful information in real-time of a surrounding context of IoMT devices, learn personalized patterns of their behaviour and change user behavior (or behavioral biases and predispositions) according to the prediction or expectation. Our Anticipatory Behavioral Process (ABP) is built on supporting the various phases of the anticipatory process as described in the following sections. 

\subsection{Three-layer architecture}
\begin{figure}
\centering
   %\begin{subfigure}[b]{0.8\textwidth}
   \includegraphics[width=1\linewidth]{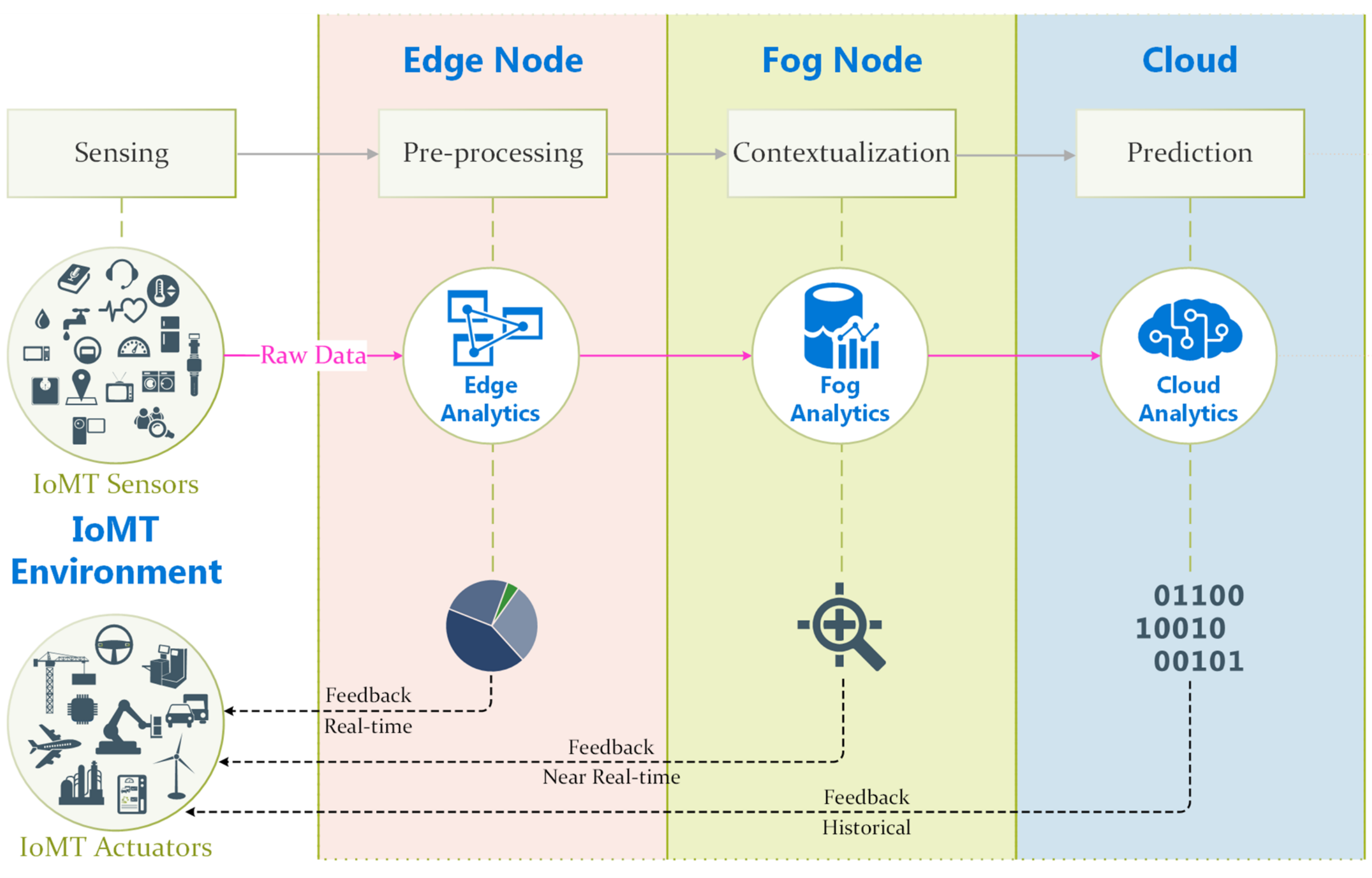}
   \caption{The Anticipatory Learning Process for IoMT in a three layer Edge - Fog - Cloud architecture}
   \label{fig:platform} 
%\end{subfigure}
\end{figure}

We map the anticipatory learning process defined above into a concrete platform composed by a hierarchy of resources: edge, fog and cloud. An overview of the proposed mapping is presented in Figure \ref{fig:platform}. Edge nodes are exploited to run the preprocessing tasks closer to IoMT sensors, rather than having the raw data transported to a data center in the cloud.  The main advantage of this approach is that edge nodes allow a faster feedback to be sent back to the IoMT actuators.  Fog nodes have the appropriate computing resources for performing the automated tasks in real-time that are needed for the contextualization phase. In data contextualization raw data streams are transformed to become suitable for learning about human mobility behaviour. Fog nodes can help improving the contextualization accuracy and reduce the computational complexity of the learning process. As the technology of IoMT sensors and actuators matures in the near future, the variety of context types will proliferate.

In the platform, IoMT devices sense data from the environment and send them to the edge nodes.  Here a preprocessing task, as described in the anticipatory model, is executed to extract information in the status of the devices which sensed the data and perform Edge Analytics such as data cleaning, basic statistics and aggregation. These analysis may help to discover, in real time, possible anomalies on the device behavior and therefore provide a fast feedback to the IoMT actuators which can take immediate actions. The analysis results with the preprocessed data streams are then transmitted to a fog node. Each fog node collects data and patterns from a group of edge nodes located in proximity of the node. Here, the additional computational power at the fog permits the data contextualization task on the received cleaned data and statistics. Through the contextualization step and fog analytics, we can learn more about the patterns and abnormalities that are happening at the edge or in general at the geographical area covered by the fog node.  The fog analytics may provide a relevant near real time feedback to the IoMT actuators that may take actions to change their behavior. The analysis results, together with the contextualized data streams, are then transmitted to the cloud node that collects all the historical network data from fog nodes and may execute complex predictive global model over the whole IoMT network. 

\subsection{Anticipatory Analytics}
One of the main contributions of the proposed platform is the definition of different kinds of anticipatory learning analytics tasks specific for each level. Table \ref{table:analytics} summarizes the main characteristics of the three architectural layers in terms of anticipatory analytics tasks. 
We first observe how the data flow tend to transmit aggregated and cleaned data from the edge to the fog and the cloud. This tend to reduce the required bandwidth as we move from the edge to the cloud. The required resources for edge computation are limited, increasing to fog up to the cloud. The data rate varies from high rates of data collected at the edge to a low rate of aggregated and cleaned data arriving at the cloud. The latency is clearly very low at the edge due to the proximity to the IoMT devices and increases as we move to the cloud. The geographical distribution is a local one at the edge as the device only sense local data while move to regional level at the fog up to the global level at the cloud. Scalability remains high at all levels since we can add as many edge and fog nodes as we need. In the cloud we can also scale infrastructure. The anticipatory learning feedback task is real time at the edge level due to the proximity to the device and the simple analytics that we can perform at this level. Feedback time increased to near real time or periodic at the fog due to the higher latency and the more complex analysis. At the cloud we can provide a periodic of historical feedback.  We also observe how the analytical complexity varies from edge to cloud. Due to the limited resource and locality of the data  the analytics at the edge may include simple data preprocessing tasks and statistics. These analysis, nevertheless, may provide a first real time feedback to the devices, like, for example detect anomalies in the device operations. The analytical complexity increases at the fog due to the increased computational resources available and the larger data available covering the fog geographical area.  Some mining algorithms can be run at this level.  Analytics at the cloud include, besides the statistics and the mining on the global network level, also the prediction task ad the consequent feedback to the IoMT actuators.

\renewcommand{\arraystretch}{1.3} 
\begin{table}
\begin{tabular} {m{5em} m{8em} m{8em} m{8em}}
\toprule
{\bf Main Characteristics} & 
{\bf Edge Analytics }& 
{\bf Fog Analytics} & 
{\bf Cloud Analytics } \\
\hline
{\bf Data Flow} & 
{\bf Input}: Raw data  {\bf Output}: Aggregated data,  Cleaned data  & 
{\bf Input}: Aggregated Data, Cleaned Data
{\bf Output}:  Contextualized Data, Aggregated data & 
{\bf Input}: Contextualized Data, Aggregated data \newline 
{\bf Output}: Predicted Values \\
%\hline
{\bf Resources} & 
Limited & 
Medium & 
High Performance (on demand) \\
%\hline
{\bf Data Rate } & 
High & 
Medium  & 
Low \\
%\hline
{\bf Latency  } & 
Low & 
Medium & 
High \\ 
%\hline
{\bf Geographical Distribution }  & 
Local  & 
Region & 
Global \\
%\hline
{\bf Scalability } & 
High & 
High & 
High \\ 
%\hline
{\bf Analytical Complexity} & 
Statistical Inference & 
Statistical Inference 
Data mining & 
Statistical Inference 
Data Mining
Machine Learning \\
%\hline
{\bf Feedback }  & 
Real-time & 
Near-real time 
Periodic & 
Periodic
Historical\\
\bottomrule
\end{tabular}
\caption{The characteristics of analytics at edge, fog and cloud}
\label{table:analytics}
\end{table}

\section{Transit Case study}
\label{sec:casestudy}

A real case study in the domain of public transit was selected to demonstrate the application of our architecture. Our case study involves the streaming of transit data of CODIAC Transit from the area of  Greater Moncton, Canada. The transit network currently operates 30 bus routes from Monday to Saturday, some of which have additional evening and Sunday services. The objective of the anticipatory learning process is to early detect anomalies in the mobility patterns and predict the punctuality of the bus services. The information related to the anomalies and the punctuality of the bus services are then used as a feedback to bus drivers, transit managers and passengers. 

The 5 phases of our anticipatory learning process are implemented using the proposed three layer architecture since each layer can support real-time, near-real time and historical feedbacks respectively. At the edge layer, the node receives the raw data stream using a 5 second time window and the WiFi access point which is available in the bus. The resulting cleaned and aggregated data from each time window are then sent to a fog node located at the Transit Operation Centre using the 3G network. The aim is to avoid storing large volumes of data at the edge  node and transport the data as soon as possible from the edge node to the fog node.  At the fog node, data is contextualized using additional information and the contextualized data sets are further sent to the cloud node which is located at our West Cloud infrastructure from Compute Canada using Ethernet connection. It is important to point out that some feedback communications have not been fully  implemented yet. In this case, we provide examples when necessary to illustrate the implementation.   

\subsection{ CODIAC Transit Feeds}
Each bus is equipped with an edge node that receives streaming transit feeds every 5 seconds that contain the GPS position and telemetry data from the sensors installed in the bus. In this experiment, the bus route 51 was selected for evaluating our anticipatory learning process because it has the highest trip density during a day. This transit data feeds consist of  a sequence $T_1$, ...$T_n$ of out-of-order tuples containing attributes in the format:

$$T_i = (S_i, xi, y_i, t_i)$$

where 

$S_i$: is a set of attributes containing telemetry data such as the bus route identifier, the bus route number, the vehicle identifier, the trip identifier, the start time of a trip, and the end time of a trip. The 17  attributes belonging to a tuple are listed in Table \ref{table:attributes};

$x_i, y_i, t_i$: are  the geographical coordinates xi,yi of the device at the sampling time $t_i$.

For the purpose of describing the outcomes from our Anticipatory Behaviour Platform, we have used 168,970 data tuples retrieved during a period of one week from 02/14/2017 to 02/20/2017. According to the transit schedule, there were 66 bus trips operated each day from Monday to Saturday and 23 bus trips on Sunday. As scheduled, each trip has taken 45 minutes.

 \begin{table}
 \small
 \centering 
\begin{tabular}{@{} m{3em} m{13em} m{15em} @{}} \toprule 
{\bf ID} & {\bf Attribute Name} &  {\bf Description }\\ \toprule
1.   &     vlr\_id & 
The ID of the data point in the vehicle location reports table.\\
2.     &  route\_id\_vlr & 
The route ID in the vehicle location reports table.\\
3.     &  route\_name & 
The route name.\\
4.    &   route\_id\_rta & 
The route ID in the route transit authority table.\\
5.    &   route\_nickname &
The abbreviation of the route.\\
6.     &  trip\_id\_br & 
The trip ID in the bid route table.\\
7.    &   transit\_authority\_service\_time\_id & 
Transit authority service time ID.\\
8.      & trip\_id\_tta&
Transit authority trip ID.\\
9.    &   trip\_start & 
Start time of the trip.\\
10.    & trip\_finish&
Finish time of the trip.\\
11.   &  vehicle\_id\_vab &
Vehicle ID.\\
12.   &  vehicle\_id\_vlr &
Vehicle ID in the vehicle locations reports table.\\
13.    & vehicle\_id\_vlr\_ta &
Descriptive name of the bus.\\
14.  &   bdescription &
Bus description.\\
15.  &   lat & 
Latitude.\\
16.   &  lng & 
Longitude.\\
17.  &    timestamp & 
Timestamp of the data point.\\
\bottomrule
\end{tabular}
\caption{The 17 attributes of the feed}
\label{table:attributes}
\end{table}

\subsection{Computation at the Edge Node}

The Cisco IR829 Industrial Integrated Services Router was used to install the edge node inside the bus. It has an Intel Atom Processor C2308 (1M Cache, 1.25 GHz) Dual Core X86 64bit, 2GB DDR3 memory and Wi-Fi connection. Our edge node h handles all routing, switching and networking traffic and a guest operating system named as runs with the IOx operating system running on a virtual machine that uses Linux Yocto. Further details are reported in \cite{Cao2017}. 
A sliding time window is generated every 5 seconds and the high volume of tuples belonging to each time window is kept in-memory until it is transported to the fog node. The tuples from the first time window are cleaned and pre-processed to remove errors and inconsistencies, and the same tasks are performed for the next time windows in a sequential manner.

\subsubsection{Data cleaning}

We implemented a Python script at the edge node which supports five automated steps to perform data cleaning. They can be described as one of the following:
\begin{enumerate}
\item Duplicated tuples: When a time window arrived at the edge node, the data tuples are sometimes transmitted twice or more. In this case, any duplicated tuple is automatically found using its timestamp and then removed. 
\item Missing tuples: Any bus trip that has at least 100 missing tuples is deleted since they may have an impact on the edge analytics results.
\item Missing attribute values:  A tuple may be corrupted and contain less than the expected number of attributes. Therefore, we fill up the missing field with “N/A” or we deleted the whole tuple depending on the kind and number of missing values. 
\item Redundant Attributes: This case happens when a new attribute is introduced to a tuple. In this case, the extra attribute is automatically deleted. 
\item Wrong attribute values: Any attribute might also contain a wrong value due to misspelling, illegal values, and uniqueness violation. In this case, the algorithm first tries to deal with the incorrect information, when not possible the value is deleted.
\end{enumerate}

In the case of our case study, 20\% of 168,970 data tuples have been deleted at the edge node,  thus making available only 137,667 tuples for data preprocessing. 

\subsubsection{Data preprocessing}

For implementing the data preprocessing phase, we have developed three steps as further described below.

\begin{figure}[!ht]
\centering
   \includegraphics[width=0.7\linewidth]{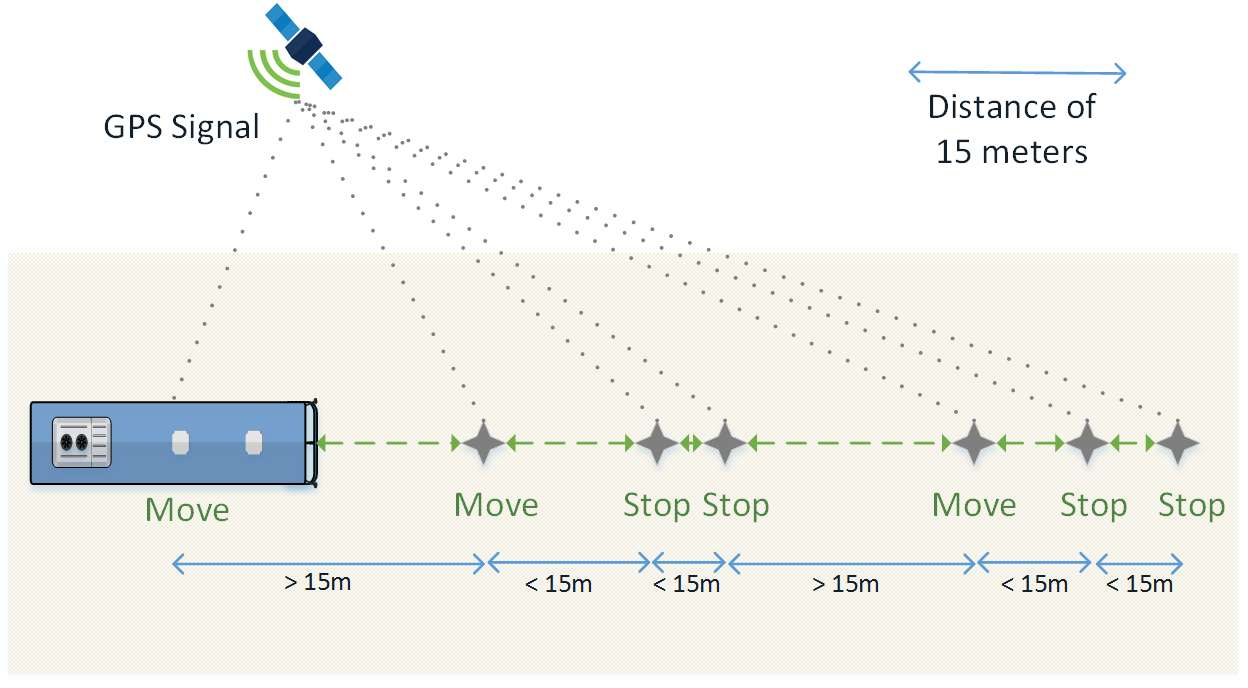}
   \caption{Moves and Stops computation}
   \label{fig:movestop} 
\end{figure}

{\bf Step 1 – Stop/Move Computation}: The aim of this step is to determine whether a bus is moving or not. The GPS coordinates of a bus position which are sent to the edge node every 5 seconds were used for this computation. In this case, a fixed distance value between two consecutive GPS positions was used for determining stops and moves (Fig. \ref{fig:movestop}) . This value was empirically determined for the CODIAC transit network as being 15 meters. When the distance between the previous point and the current point is more than 15m, the bus is moving therefore the current point is tagged as a move. In contrast, when the distance is less than 15m, the current point is tagged as a stop.

{\bf Step 2 – Temporal Aggregation}:This step is performed at the end of each trip and it is used to compute (i) the actual time duration of a trip using the timestamps of the origin and destination points of each trip, (ii) the total number of stops during a trip , and (iii) the total number of moves during a trip. In summary, five data fields 
 (Trip Id, Date, Start\_Time, Total\_Move, Total\_Stop, Total\_Time\_Length) were used for the temporal computations. 

{\bf Step 3 – Summary Function}: In this step we computed the average trip time in the morning (5h-12h), afternoon (13h-18h), and evening (19h-24h). Besides, the average of the total number of moves and stops was computed for the different times of the day (i.e. morning, afternoon, evening). 
In the case of our case study, 137,667 tuples have been used in the data preprocessing phase and as a result, they became ready to be used for edge analytics and be later transported to the fog node.

\subsubsection{Anticipatory Edge Analytics}

\begin{figure}[!ht]
\centering
   %\begin{subfigure}[b]{0.8\textwidth}
   \includegraphics[width=0.9\linewidth]{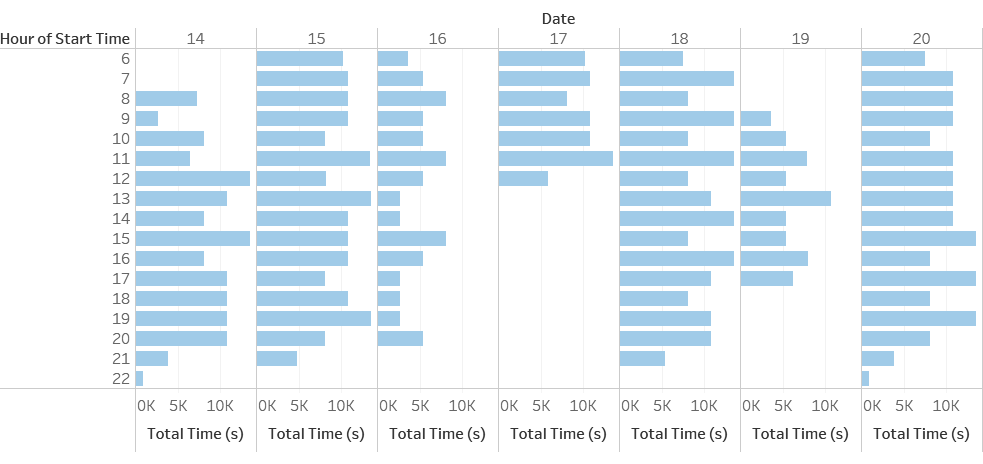}
   \caption{Overview of the hourly trip times for each day of the week}
   \label{fig:plot1} 
%\end{subfigure}
\end{figure}

The aim of implementing the anticipatory analytics was to demonstrate how it is possible to learn about real-time abnormalities in the transit network, for example, learning about the interruption of services.  Fig. \ref{fig:plot1} illustrates the existence of several missing trips that have been detected in real-time. The buses did not run on February 14th at 6h to 7h; and there were no trips at 22h on the 15th, 16th, 18th. Moreover, missing trips have also occurred on the 17th after 12h and on the 19th early in the morning (6h and 7h) and in the evening (18h to 22h). This is relevant information for the real-time feedback because it can generate warnings to the transit managers as well as passengers about the current state of the network at the bus line level.
Moreover, computing the total trip time in real-time can provide relevant information to the transit manager about the abnormalities occurring with the bus services. For example, Figure \ref{fig:plot1} shows the total trip times from February 14th to February 20th.  On February 14th, the shortest trip has lasted for 897 seconds (at 22h of the start time), meanwhile the longest trip has taken 13,468 seconds (at 12h of the start time). The weather conditions were fair on that day, making such an information relevant as a feedback to the transit manager in order to identify the actual cause of these disruptions on the bus service. In contrast, on February 16th due to a snowstorm the bus service was erratic as shown by the different values of the total trips. This information is relevant as a feedback to be provided to the passengers in such a way that they would be able to make a decision of taking a bus or search for another mode of transportation.

\subsection{Computation at the Fog}
The fog node was implemented using the Cisco UCS 240 modular with two rack-unit (2RU) server and 2 Intel Xeon processor E5-2600 CPUs, 24 double-data-rate 4 (DDR4) dual in-line memory (DIMMs) of up to 2400 MHz speeds, 6 PCI Express (PCIe) Generation 3 slots, 12 large-form factor hard drives. The dataflow is managed using Cisco Kinetics functionalities such as Broker, Protocol and DSL links, and Data Control management.   
Every 6 hours, all tuples were scheduled to be sent to the fog node located at the Transit Operation Centre. This is implemented using the Message Broker available at the fog node. On February 20th, 137,667 tuples and summaries were transmitted from the edge to the fog node. 
At the fog node we have the capability to handle the regional geo-distribution of data and we have enough computing resources to perform the data contextualization phase and fog analytics.  All necessary information including the GTFS (General Transit Feed Specification) data [], GIS layers, the geographical location of all the bus stations and the PostgreSQL database are available at the fog node. It is worth noticing that a single fog node can handle data being pushed from more than one edge node, therefore, several trips from different bus routes can be contextualized and analyzed at the same time.  

\subsubsection{Data Contextualization}
\begin{figure}[!ht]
\centering
   %\begin{subfigure}[b]{0.8\textwidth}
   \includegraphics[width=1\linewidth]{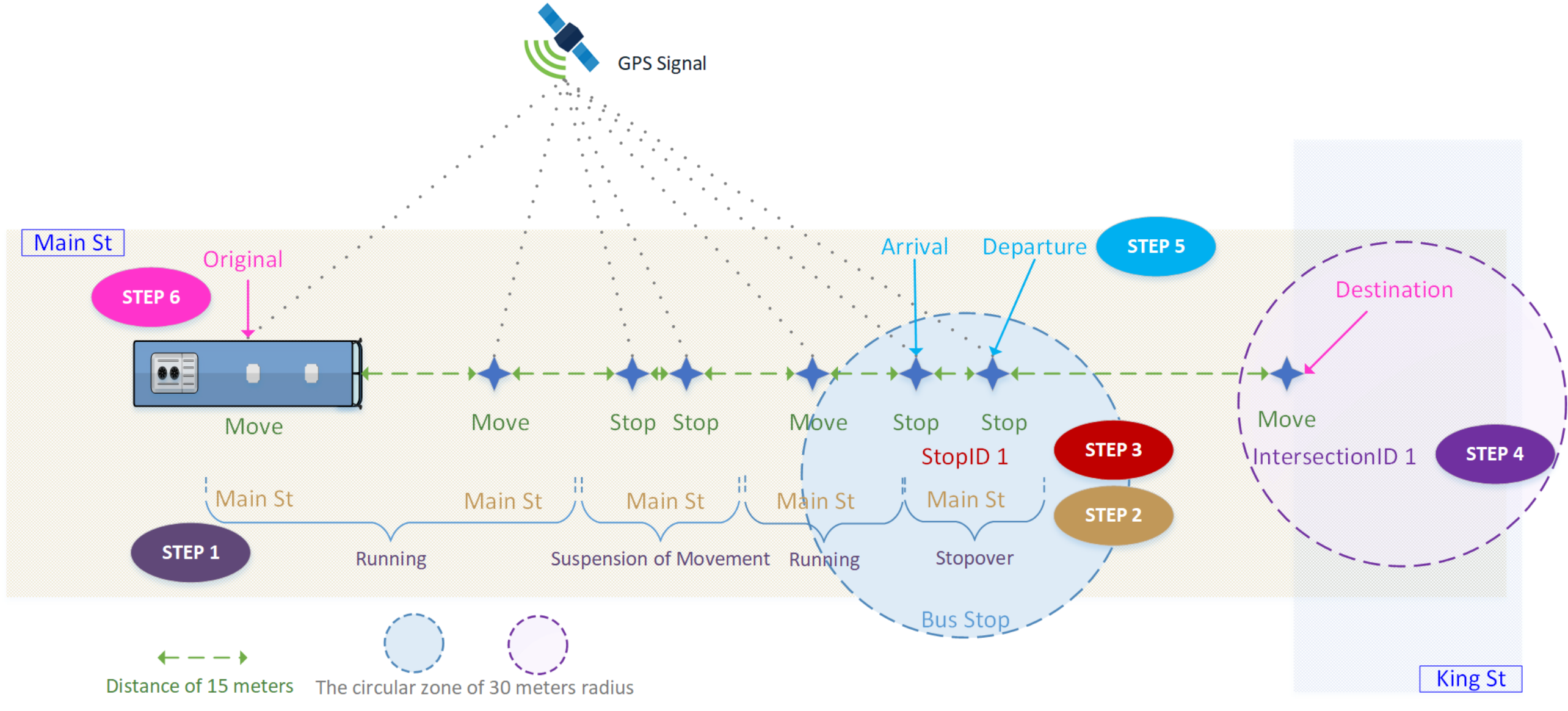}
   \caption{Example of the stops/moves contextualization}
   \label{fig:stopmove} 
%\end{subfigure}
\end{figure}

The implementation of the data contextualization phase is illustrated using 6 steps and illustrated in Figure \ref{fig:stopmove}. We would like to emphasize the fact that all these steps are fully automated and do not require any human intervention.  

{\bf Step 1: Categorizing Stop/Move}. The aim of this step is to classify the moves and stops obtained from the edge node to improve our understanding about their context. Any stop may occur due to a traffic jam, accident, by collecting large number of passengers at a bus station, or waiting at a traffic light. In contrast, moves might happen when a bus is moving on a street or intersection, and passing in front of a bus station. Therefore, the classification of stops and moves is carried out by adding one new attribute (label) as specified in Table \ref{table:stopmove}.

\begin{table}
 \small
 \centering 
\begin{tabular}{@{} m{5em} m{6em} m{20em} @{}} \toprule 
{\bf Stop/Move} & {\bf Label } & {\bf Description}\\ \toprule
Move &  Running & When a bus is moving on a street\\
Move & Passing & When a bus passes in front of a bus station without stopping, for example when there are no passengers to drop off or get on \\
Stop & Suspension & It may occur due to an intersection, stop sign, accident, or traffic jam.\\
Stop& Stopover & When a bus stops at a bus station for dropping off or picking up passengers\\
\bottomrule
\end{tabular}
\caption{Stop and move classification}
\label{table:stopmove}
\end{table}

The classification algorithm works by creating a circular buffer with a radius of 30m (estimated empirically) for each bus station, and geographically matching it with a stop or move (i.e. stopover and passing) location of a bus. The stops which are located inside the buffer were classified as “stopovers”, otherwise they were classified as “suspension of movement”. Moreover, the moves which were located inside the buffer are classified as “passing”, otherwise they were classified as “running” on a street.

{\bf Step 2: Street Name Annotation}. The objective of this step is to annotate the moves and stops according to the streets nomenclature of a transit network. For this step, a query on PostgreSQL database is run to retrieve the names of the street where a move or stop is located. This is a non-trivial step because the geographical coordinates of the stops and moves are obtained from GPS signals which normally have 10m of accuracy in urban areas \cite{salarian2015accurate}.

{\bf Step 3: Bus Station Identification}. The objective of this contextualization step is to associate a bus station identifier to each bus movement tuple labeled as stopover and passing. This is a crucial step to provide a link between the movement data to the bus station information available from the GTFS data. It is important to point out that the algorithm also needs to verify the direction of a moving bus (e.g. eastbound and westbound) in order to identify the bus station that a stopover/passing is actually located. To do so, we select the tuple located at the middle of a bus route for using it as a reference point for identifying the direction of a moving bus. Each stop is then annotated as “outbound” and “return” values.

{\bf Step 4: Street Intersection Identification}. Here we tag street intersections to each tuple. The algorithm creates a circular zone with a radius of 30m (determined empirically) for each street intersection. The tuples containing stops and moves that are located inside the circular zone are tagged with the intersection identifier. 

{\bf Step 5: Arrival/Departure Times Identification}.  The aim of this step is to determine the actual arrival and departure time of a bus at a stopover for dropping off or picking up passengers. Our algorithm computes the timestamp of the first stop point in a stopover within the circular zone of 30m radius (determined empirically) around each bus station, and considers it as the actual arrival time. Similarly, the timestamp of the last point of a stopover point within a circular buffer from the same station is considered the departure time. 

{\bf Step 6: Origin/Destination Trip Identification}. During this step we identify origin and destination of a trip. We tag each first tuple of a bus trip as origin, and each last tuple of a bus trip as destination. The remaining tuples are then sequentially indexed.
After finishing the contextualization steps, 6 new data fields were generated and attached to the cleaned data tuples which were received from the edge node. The preprocessed data streams are now transformed to contextualized data streams every 6 hours and ready for the fog analytics.

\subsubsection{ Fog Analytics}

\begin{figure}[!ht]
\centering
   %\begin{subfigure}[b]{0.8\textwidth}
   \includegraphics[width=0.8\linewidth]{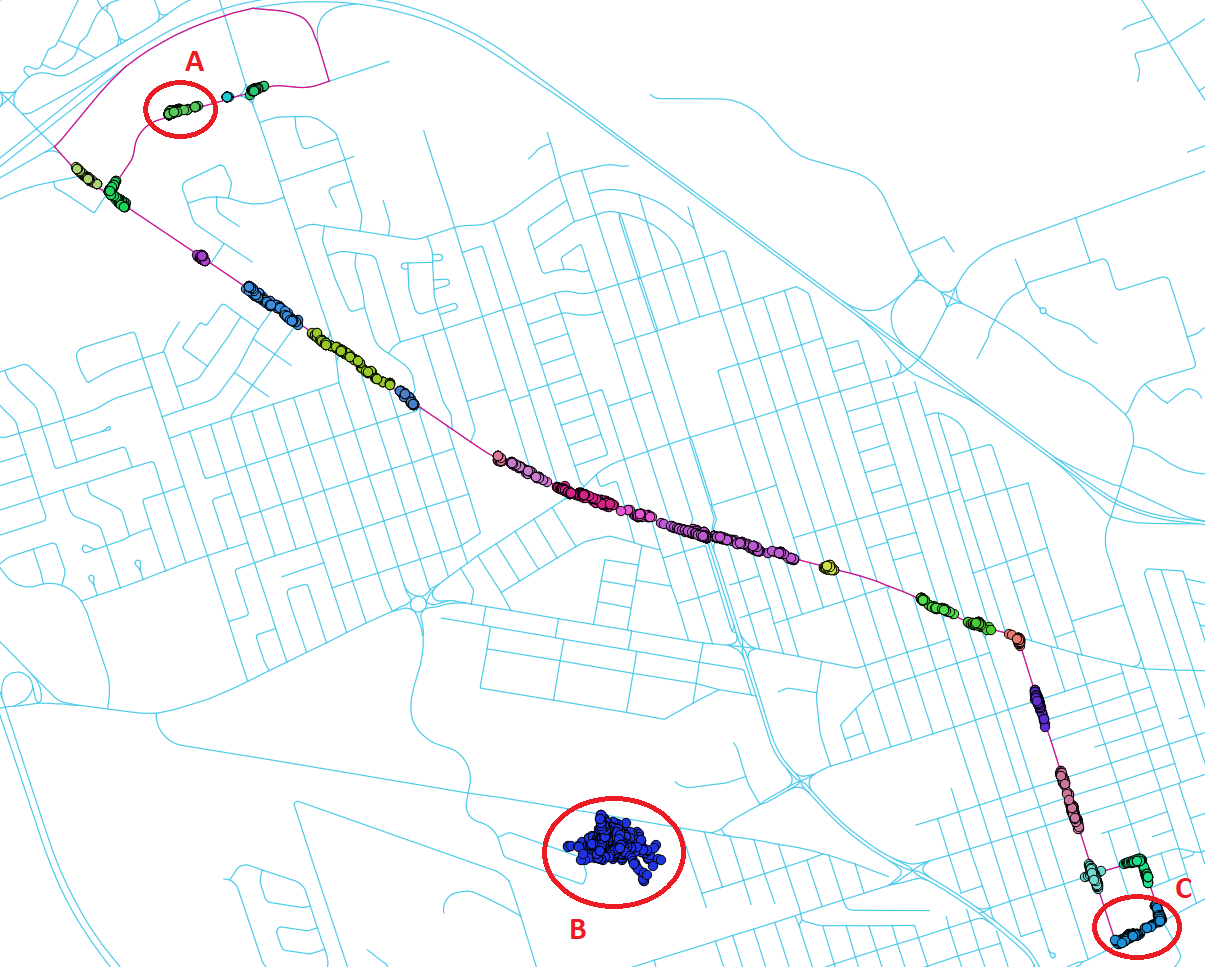}
   \caption{Overview of the spatial clusters that were computed at the fog node.}
   \label{fig:cluster} 
%\end{subfigure}
\end{figure}

The DBSCAN algorithm cite{Ester1996dbscan} was selected to detect spatial clusters, considering as parameters a spatial radius of 15 meters and the minimum number of points in a cluster of 8. We used the Skitlearn DBSCAN algorithm  and applied it to all the stops every 6 hours with the purpose of discovering spatial clusters which might indicate traffic congestions. The spatial clusters contain in average up to 201 stops. In  Figure \ref{fig:cluster}, we show an example on February 14th when 24 spatial clusters were found, and many of them were located along the streets and the Transit Operational Centre in Moncton. This information is relevant to be provided as a feedback to the transit managers as soon as it was obtained because they represent disruptions on the bus service due to traffic congestions such as  cluster A  (2766 stops) that is located at the Plaza Blvd bus station (Bus Stop ID 6810785) and cluster C (1894 stops) that is located at the Main St bus station (Bus Stop ID 6810785). It is also important to point out that cluster B (2327 stops)  that is located at  the Transit Operation Centre reinforce the information obtained from the edge node that the bus service was interrupted due to the snowstorm on this day.  

Once the spatial clusters were computed, all tuples were stored permanently in the PostgreSQL database.  Only the tuples found in the spatial clusters were finally transmitted to our cloud environment.

\subsection{Computation at the Cloud} 
At the cloud cluster, our Anticipatory Behaviour Platform is implemented with the support of Compute Canada which provides an IaaS where we have created and allocated cloud resources such as VMs, Servers, Storages, Load Balancers, IP addresses. Our cloud capabilities include maximum 5 Instances, 40 VCPUs, 150GB RAM, 2 Floating IPs, 5TB Volume Storage. 

Once all tuples belonging to all spatial clusters  have arrived at the cloud (i.e. 30,746), a HDFS file is generated and used for predicting the punctuality of the bus services from a network perspective. In order to achieve that, a random forest algorithm was used to build a predictive model. Our purpose here is to illustrate the potential of using our predictive model for classifying the new spatial clusters. We have not analyzed the deterioration of our predictive model. It is also important to point out that the prediction here is only to illustrate the role of our  three layer architecture in supporting an anticipatory learning process, rather than finding the best punctuality prediction model. 
In summary, we have assigned the target label for each data tuple in the training dataset based on 3 categories described as one of the following:

\begin{itemize}
\item[--] On time: this label indicates whether the bus is expected to arrive on time at a specific bus station; 
\item[--] Early: this label indicates whether the bus is expected to arrive earlier at the bus station;
\item[--] Late: this label indicates whether the bus is expected to arrive later at the bus station.
\end{itemize}

For the implementation, we have used the range of 1.3 minutes (80s) for the early label and  5.3 minutes (320s) for the late label as suggested by \cite{diab2015transit}.  The time computation was based on the timestamps of the tuples and the timestamps of the scheduled arrival times obtained from the GTFS data.

\begin{table}
 \small
 \centering 
\begin{tabular}{@{} m{10em} m{15em} @{}} \toprule 
{\bf Attribute} & {\bf Description  } \\ \toprule 
trip\_id & this is the identifier of the trip\\
lat  &  the latitude \\
lng &  the longitude\\
gps\_timestamp & the timestamp \\
street\_name & the name of the street \\
direction & the direction of the bus movement\\
stop\_id & the identifier of the stop \\
movement\_sequence & the movement sequence of the bus\\
arrival\_time & the arrival at the stop \\
target\_class & early/on time/late\\
\bottomrule
\end{tabular}
\caption{ List of the most influential attributes in the prediction model}
\label{table:relevant}
\end{table}

\begin{figure}[!ht]
\centering
   \includegraphics[width=0.5\linewidth]{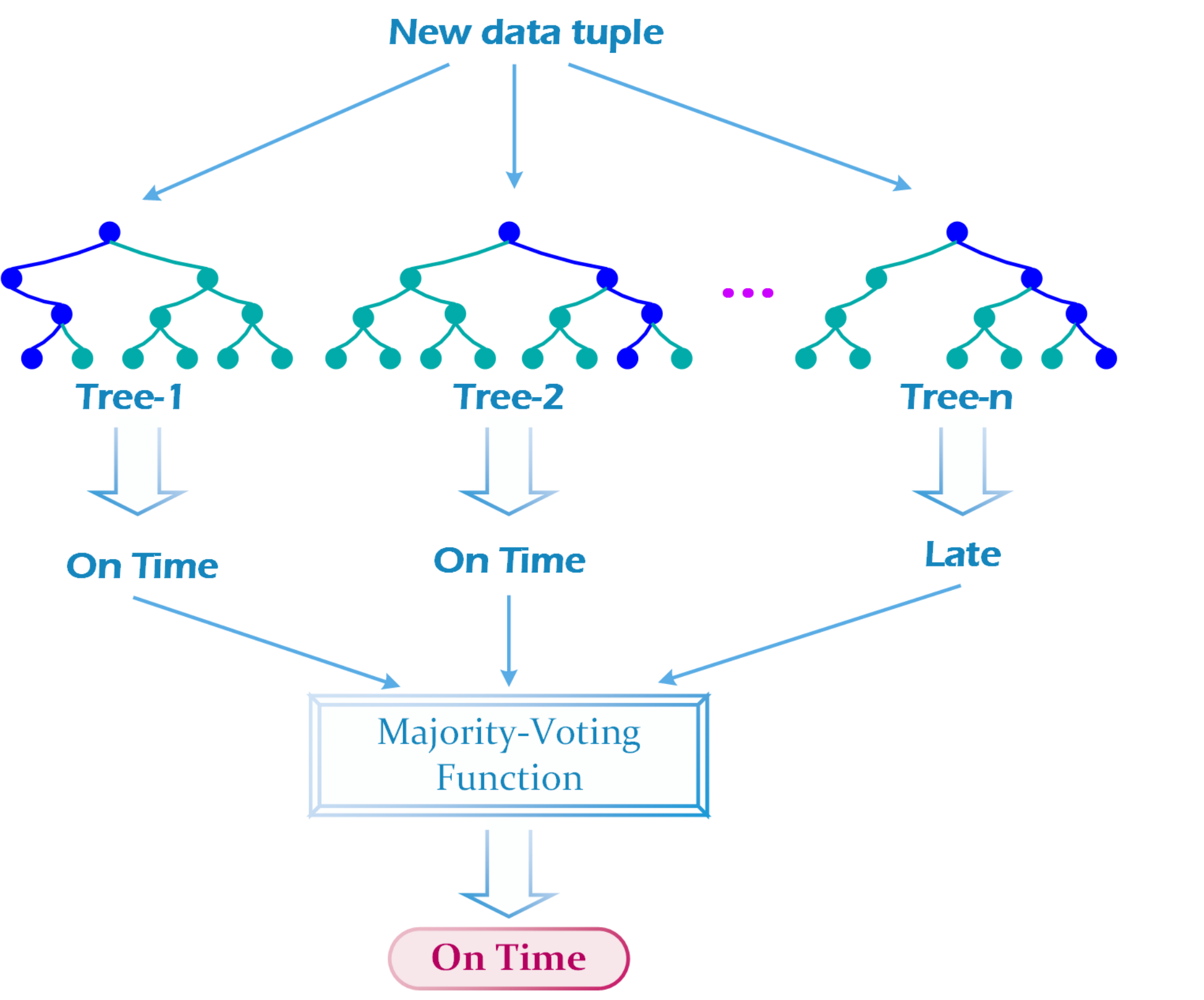}
   \caption{Random Forest model with majority voting}
   \label{fig:randomforest} 
\end{figure}

Figure \ref{fig:randomforest} ddepicts the predictive model showing a number of decision trees that were created during the training phase. Each decision tree contains a random subset of the 9 most relevant attributes. When a new data tuple comes to the prediction model, it is predicted through each decision tree and returns the target class label. A majority-voting function was utilized to vote the majority target class label and predict the label. 

\begin{figure}[!ht]
\centering
   \includegraphics[width=0.8\linewidth]{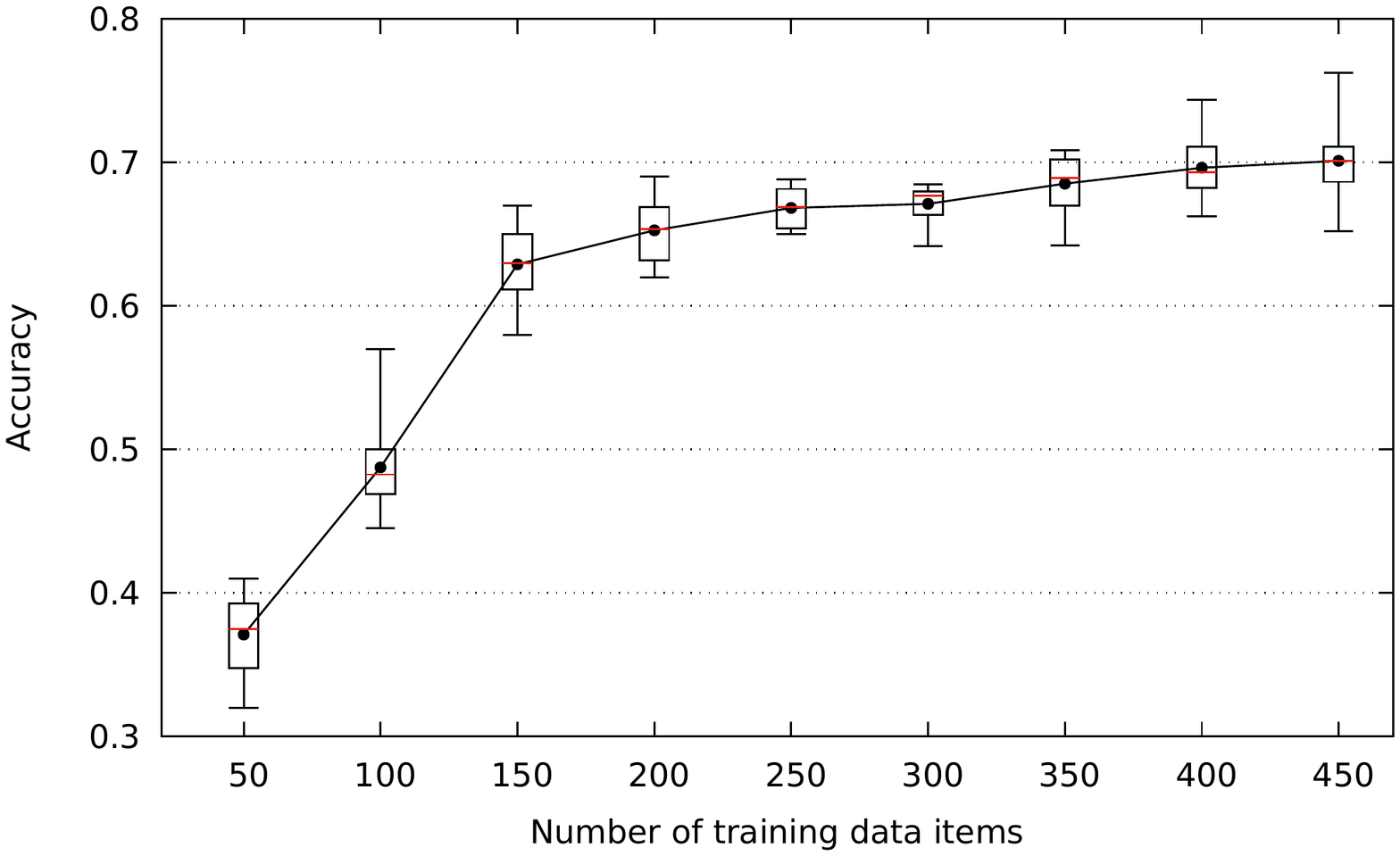}
   \caption{Accuracy of the prediction based on number of training items}
   \label{fig:candlestick} 
\end{figure}

We evaluated our predictive model applying 10-fold cross validation by partitioning the data into a training set to train the model, and a test set to evaluate it.  We then computed the average accuracy of the model. To evaluate how does the accuracy of the prediction model changes as a function of the training-set size, we have plot the accuracy curve as shown in Figure \ref{fig:candlestick} . This plot indicates that, not surprisingly, when training data samples increase, the accuracy of our predictive model increases. Other machine learning algorithms should be tested (e.g. gradient boost)  and compared with state of the art to find the best prediction performance.

At the end of the computation at the cloud, the predicted values become the historical feedback that can go back to the transit manager in order to understand the how efficient the bus service is at the transit network level during a long period of time. In this experiment we have only used the data generated by one bus, however, the the platform can be applied to the whole transit network.

\section{Conclusions and Future Work}
\label{sec:conclusions}
This paper presents the conceptual design and overall description of a multi-layered edge-fog-cloud architecture targeting an anticipatory learning process in an IoMT domain. The Anticipatory Behavioral Platform meets a concept that is gaining more and more momentum in the way analytics of data is performed: whenever possible push the computation toward the edge while trying to keep the data as close as possible from where they originated. This presents immediate general advantages that would be favourable to almost any IoMT application: it guarantees data privacy (at least to a certain extent), reduces the cost to transfer large amount of data toward datacenters, and makes possible to receive fast feedback by the analysis of the data. In particular this last point is of fundamental importance for the real-time anticipatory learning processes we target in the paper.

Beside its description, we also have experimented the proposed architecture on an actual real-world case study for the management of public transit in the Moncton area in Canada.  The case study illustrates the usefulness and the effectiveness of the proposed architecture.

In fact, our lesson learned is that each layer of the architecture considered in isolation would not be able to manage the anticipatory learning process. 
%Sure, the combination requires effort into coordinating the resources and possibly reworking several of the data analytics process involved in the anticipatory learning, but it worths the hassle. 
As a matter of fact, realising a complete anticipatory learning process using only edge devices would be not feasible due to the lack of computing power and storage on such devices, and would have required an extreme simplification of the prediction and analytics processes.
Still, using a combination of fog and edge resources would not scale when increasing the number of monitored buses and lines. Finally, using only the cloud would have made the latency of the feedback unacceptable, still it is a useful resources that provides (the illusion of) infinite scalability for batch processing and complex data analysis, while having the convenience of using the resources (and pay for them, in case of public clouds) only when needed. 

For future research work, we plan to extend the architecture into those scenarios in which the direct communication between nodes belonging to the same layers is needed. This would open new challenges and opportunities, since that would be possible to use other peers information to improve the knowledge on the whole system locally at each node, improving the overall precision of the prediction mechanisms and, as consequence, the effectiveness of the anticipatory learning process.

\section*{Acknowledgments}
This work was supported by YYYYYY.

%\section*{References}
\bibliographystyle{elsarticle-num}
\bibliography{biblio}

\begin{thebibliography}{10}
\expandafter\ifx\csname url\endcsname\relax
  \def\url#1{\texttt{#1}}\fi
\expandafter\ifx\csname urlprefix\endcsname\relax\def\urlprefix{URL }\fi
\expandafter\ifx\csname href\endcsname\relax
  \def\href#1#2{#2} \def\path#1{#1}\fi

\bibitem{lopez2012adding}
T.~S. L{\'o}pez, D.~C. Ranasinghe, M.~Harrison, D.~McFarlane, Adding sense to
  the internet of things, Personal and Ubiquitous Computing 16~(3) (2012)
  291--308.

\bibitem{gerla2014internet}
M.~Gerla, E.-K. Lee, G.~Pau, U.~Lee, Internet of vehicles: From intelligent
  grid to autonomous cars and vehicular clouds, in: Internet of Things
  (WF-IoT), 2014 IEEE World Forum on, IEEE, 2014, pp. 241--246.

\bibitem{lu2014connected}
N.~Lu, N.~Cheng, N.~Zhang, X.~Shen, J.~W. Mark, Connected vehicles: Solutions
  and challenges, IEEE internet of things journal 1~(4) (2014) 289--299.

\bibitem{rosen2012anticipatory}
R.~Rosen, Anticipatory systems, in: Anticipatory systems, Springer, 2012, pp.
  313--370.

\bibitem{nadin2010anticipatory}
M.~Nadin, Anticipatory computing. from a high-level theory to hybrid computing
  implementations, International journal of applied research on information
  technology and computing (IJARITAC) 1~(1) (2010) 1--27.

\bibitem{pejovic2015anticipatory}
V.~Pejovic, M.~Musolesi, Anticipatory mobile computing: A survey of the state
  of the art and research challenges, ACM Computing Surveys (CSUR) 47~(3)
  (2015) 47.

\bibitem{stojmenovic2014fog}
I.~Stojmenovic, S.~Wen, The fog computing paradigm: Scenarios and security
  issues, in: Computer Science and Information Systems (FedCSIS), 2014
  Federated Conference on, IEEE, 2014, pp. 1--8.

\bibitem{li2015internet}
S.~Li, L.~Da~Xu, S.~Zhao, The internet of things: a survey, Information Systems
  Frontiers 17~(2) (2015) 243--259.

\bibitem{chen2014big}
M.~Chen, S.~Mao, Y.~Liu, Big data: A survey, Mobile Networks and Applications
  19~(2) (2014) 171--209.

\bibitem{shi2016edge}
W.~Shi, J.~Cao, Q.~Zhang, Y.~Li, L.~Xu, Edge computing: Vision and challenges,
  IEEE Internet of Things Journal 3~(5) (2016) 637--646.

\bibitem{intel2014}
Nokia, Intel, {Increasing Mobile Operators Value Proposition With Edge
  Computing}, \url{https://www.intel.co.id/content
  /dam/www/public/us/en/documents/technology-briefs
  /edge-computing-tech-brief.pdf}, [Online; accessed 15-November-2017] (2014).

\bibitem{hu2015mobile}
Y.~C. Hu, M.~Patel, D.~Sabella, N.~Sprecher, V.~Young, Mobile edge
  computing—a key technology towards 5g, ETSI White Paper 11~(11) (2015)
  1--16.

\bibitem{bonomi2012fog}
F.~Bonomi, R.~Milito, J.~Zhu, S.~Addepalli, Fog computing and its role in the
  internet of things, in: Proceedings of the first edition of the MCC workshop
  on Mobile cloud computing, ACM, 2012, pp. 13--16.

\bibitem{satyanarayanan2009case}
M.~Satyanarayanan, P.~Bahl, R.~Caceres, N.~Davies, The case for vm-based
  cloudlets in mobile computing, IEEE pervasive Computing 8~(4).

\bibitem{khan2013towards}
A.~N. Khan, M.~M. Kiah, S.~U. Khan, S.~A. Madani, Towards secure mobile cloud
  computing: A survey, Future Generation Computer Systems 29~(5) (2013)
  1278--1299.

\bibitem{bonomi2014fog}
F.~Bonomi, R.~Milito, P.~Natarajan, J.~Zhu, Fog computing: A platform for
  internet of things and analytics, in: Big Data and Internet of Things: A
  Roadmap for Smart Environments, Springer, 2014, pp. 169--186.

\bibitem{yi2015survey}
S.~Yi, C.~Li, Q.~Li, A survey of fog computing: concepts, applications and
  issues, in: Proceedings of the 2015 Workshop on Mobile Big Data, ACM, 2015,
  pp. 37--42.

\bibitem{ali2009green}
M.~Ali, Green cloud on the horizon, Cloud Computing (2009) 451--459.

\bibitem{kumar2010cloud}
K.~Kumar, Y.-H. Lu, Cloud computing for mobile users: Can offloading
  computation save energy?, Computer 43~(4) (2010) 51--56.

\bibitem{banerjee2011everything}
P.~Banerjee, R.~Friedrich, C.~Bash, P.~Goldsack, B.~Huberman, J.~Manley,
  C.~Patel, P.~Ranganathan, A.~Veitch, Everything as a service: Powering the
  new information economy, Computer 44~(3) (2011) 36--43.

\bibitem{hong2013mobile}
K.~Hong, D.~Lillethun, U.~Ramachandran, B.~Ottenw{\"a}lder, B.~Koldehofe,
  Mobile fog: A programming model for large-scale applications on the internet
  of things, in: Proceedings of the second ACM SIGCOMM workshop on Mobile cloud
  computing, ACM, 2013, pp. 15--20.

\bibitem{atzmueller2016sensing}
M.~Atzmueller, B.~Fries, N.~Hayat, Sensing, processing and analytics:
  augmenting the u bicon platform for anticipatory ubiquitous computing, in:
  Proceedings of the 2016 ACM International Joint Conference on Pervasive and
  Ubiquitous Computing: Adjunct, ACM, 2016, pp. 1239--1246.

\bibitem{nahrstedt2016internet}
K.~Nahrstedt, H.~Li, P.~Nguyen, S.~Chang, L.~Vu, Internet of mobile things:
  Mobility-driven challenges, designs and implementations, in:
  Internet-of-Things Design and Implementation (IoTDI), 2016 IEEE First
  International Conference on, IEEE, 2016, pp. 25--36.

\bibitem{sun2016moving}
W.~Sun, J.~Zhu, N.~Duan, P.~Gao, G.~Q. Hu, W.~S. Dong, Z.~H. Wang, X.~Zhang,
  P.~Ji, C.~Y. Ma, et~al., Moving object map analytics: A framework enabling
  contextual spatial-temporal analytics of internet of things applications, in:
  Service Operations and Logistics, and Informatics (SOLI), 2016 IEEE
  International Conference on, IEEE, 2016, pp. 101--106.

\bibitem{vieira2015usapiens}
M.~R. Vieira, L.~Barbosa, M.~Korm{\'a}ksson, B.~Zadrozny, Usapiens: A system
  for urban trajectory data analytics, in: Mobile Data Management (MDM), 2015
  16th IEEE International Conference on, Vol.~1, IEEE, 2015, pp. 255--262.

\bibitem{herrera2016smart}
L.~F. Herrera-Quintero, K.~Banse, J.~Vega-Alfonso, A.~Venegas-Sanchez, Smart
  its sensor for the transportation planning using the iot and bigdata
  approaches to produce its cloud services, in: Telematics and Information
  Systems (EATIS), 2016 8th Euro American Conference on, IEEE, 2016, pp. 1--7.

\bibitem{welbourne2009building}
E.~Welbourne, L.~Battle, G.~Cole, K.~Gould, K.~Rector, S.~Raymer,
  M.~Balazinska, G.~Borriello, Building the internet of things using rfid: the
  rfid ecosystem experience, IEEE Internet computing 13~(3).

\bibitem{somov2013supporting}
A.~Somov, C.~Dupont, R.~Giaffreda, Supporting smart-city mobility with
  cognitive internet of things, in: Future Network and Mobile Summit
  (FutureNetworkSummit), 2013, IEEE, 2013, pp. 1--10.

\bibitem{wang2012walksafe}
T.~Wang, G.~Cardone, A.~Corradi, L.~Torresani, A.~T. Campbell, Walksafe: a
  pedestrian safety app for mobile phone users who walk and talk while crossing
  roads, in: Proceedings of the Twelfth Workshop on Mobile Computing Systems \&
  Applications, ACM, 2012, p.~5.

\bibitem{krause2006context}
A.~Krause, A.~Smailagic, D.~P. Siewiorek, Context-aware mobile computing:
  Learning context-dependent personal preferences from a wearable sensor array,
  IEEE Transactions on Mobile Computing 5~(2) (2006) 113--127.

\bibitem{larose2014discovering}
D.~T. Larose, Discovering knowledge in data: an introduction to data mining,
  John Wiley \& Sons, 2014.

\bibitem{saez2016inffc}
J.~A. S{\'a}ez, M.~Galar, J.~Luengo, F.~Herrera, Inffc: an iterative class
  noise filter based on the fusion of classifiers with noise sensitivity
  control, Information Fusion 27 (2016) 19--32.

\bibitem{barnaghi2013data}
P.~Barnaghi, A.~Sheth, C.~Henson, From data to actionable knowledge: Big data
  challenges in the web of things [guest editors' introduction], IEEE
  Intelligent Systems 28~(6) (2013) 6--11.

\bibitem{kuhn2013applied}
M.~Kuhn, K.~Johnson, Applied predictive modeling, Vol. 810, Springer, 2013.

\bibitem{perera2014context}
C.~Perera, A.~Zaslavsky, P.~Christen, D.~Georgakopoulos, Context aware
  computing for the internet of things: A survey, IEEE Communications Surveys
  \& Tutorials 16~(1) (2014) 414--454.

\bibitem{van2005context}
A.~H. Van~Bunningen, L.~Feng, P.~M. Apers, Context for ubiquitous data
  management, in: Ubiquitous Data Management, 2005. UDM 2005. International
  Workshop on, IEEE, 2005, pp. 17--24.

\bibitem{abowd1999towards}
G.~Abowd, A.~Dey, P.~Brown, N.~Davies, M.~Smith, P.~Steggles, Towards a better
  understanding of context and context-awareness, in: Handheld and ubiquitous
  computing, Springer, 1999, pp. 304--307.

\bibitem{henricksen2003framework}
K.~Henricksen, A framework for context-aware pervasive computing applications,
  University of Queensland Queensland, 2003.

\bibitem{Cao2017}
H.~Cao, M.~Wachowicz, \href{https://arxiv.org/abs/1705.08449v2}{Developing an
  edge computing platform for real-time descriptive analytics}.
\newline\urlprefix\url{https://arxiv.org/abs/1705.08449v2}

\bibitem{salarian2015accurate}
M.~Salarian, A.~Manavella, R.~Ansari, Accurate localization in dense urban area
  using google street view images, in: SAI Intelligent Systems Conference
  (IntelliSys), 2015, IEEE, 2015, pp. 485--490.

\bibitem{diab2015transit}
E.~Diab, M.~Badami, A.~El‐Geneidy, Bus transit service reliability and
  improvement strategies: Integrating the perspectives of passengers and
  transit agencies in north america, Transport Reviews 23~(3) (2015) 292 –
  328.

\end{thebibliography}
\end{document}